\documentclass[12pt]{article}

\usepackage{a4wide,graphics,graphicx,amsmath,amssymb,cite,nicefrac}
\usepackage[verbose]{wrapfig}
\usepackage{authblk,hyperref}
\catcode`@=11 \@addtoreset{equation}{section} \catcode`@=12

\begin{document}
\date{}
\title{
{\baselineskip -.2in
%\vbox{\small\hskip 4in \hbox{hep-th/07-10~~~~~~~~~~}}
%\vbox{\small\hskip 4in \hbox{IITM/PH/TH/2014/2}}
} 
\vskip .4in
\vbox{
{\bf \LARGE On the Uniqueness of  Supersymmetric Attractors}
}}
%\narrowtext
%\author{Samrat Bhowmick\thanks{email: samrat@physics.iitm.ac.in}}
\author{Taniya Mandal\thanks{email: taniya@physics.iitm.ac.in} } 
\author{Prasanta K. Tripathy\thanks{email: prasanta@iitm.ac.in}}
\affil{\normalsize\it Department of Physics, \authorcr \it Indian Institute of Technology Madras, \authorcr \it Chennai 600036, India.}
\maketitle
\begin{abstract}

In this paper we discuss the uniqueness of supersymmetric attractors in four dimensional $N=2$ supergravity theories
coupled to $n$ vector multiplets. We prove that for a given charge configuration the supersymmetry preserving axion 
free attractors are unique. We generalise the analysis to axionic attractors and state the conditions for uniqueness 
explicitly. We consider the example of a two-parameter model and find all solutions to the supersymmetric attractor
equations and discuss their uniqueness. 
\end{abstract}

\newpage
\section{Introduction}

Understanding the origin of black hole entropy has  remained to be an important topic of research in gravity 
and string theory since the seminal work by Bekenstein\cite{Bekenstein:1973ur} on this issue. One of the important 
developments in this area is 
the so called attractor mechanism, which states that, in a theory of gravity coupled to several scalar fields admitting a 
single centred extremal black hole, the scalar fields run into a fixed point at the horizon whose value depends only 
on the black hole charges\cite{Ferrara:1995ih,Strominger:1996kf,Ferrara:1997tw,Goldstein:2005hq}. There are several 
aspects of attractor mechanism which have been studied thoroughly \cite{Ferrara:2008hwa,Bellucci:2007ds}. 
Multiplicity of the attractors  is one of the puzzling issues which remains to be understood better. Because of the 
presence of multiple basin of attractors, the near horizon geometry of the black hole is no longer uniquely determined 
by its charges and one needs to specify the area code in addition to the black hole charges.%\cite{Moore:1998pn,Moore:1998zu}. 

The existence of multiple basin of attractors for a given set of charges has been first discussed in 
\cite{Moore:1998pn,Moore:1998zu}. Area codes in the context of flux vacua and black hole attractors has been
studied  \cite{Giryavets:2005nf,Misra:2007yu}. Subsequently, multiple supersymmetric attractors in five dimensional 
$N=2$ supergravity theory has been discussed and explicit constructions in the simple case of a two parameter 
model has been carried out \cite{Kallosh:1999mz}. The analysis
has been extended to four dimensional $N=2$ supergravity \cite{Dominic:2014zia} by using the know $4D-5D$ 
correspondence of the attractor points \cite{Ceresole:2007rq}. Further, new multiple non-supersymmetric attractors 
which does not have obvious five dimensional embedding has been constructed \cite{Dominic:2014zia}. 
Multiple attractors in a one parameter model in the presence of quantum corrections has already been studied
 \cite{Bellucci:2007eh}.

The existence of multiple single centred supersymmetric attractors might at first sight appear to be in contradiction 
with the uniqueness results\cite{Wijnholt:1999vk}. (For  homogeneous moduli spaces, the solution is 
always unique up to a duality transformation \cite{Ceresole:1995jg}). However, as explained by Kallosh
\cite{Kallosh:1999mb}, this is not always the case, because the moduli 
space might in general possess several disconnected branches. The attractor solution in each of these branches
remains unique. One might expect similar results in four dimensional $N=2$ supergravity. However, though there 
exists multiple non-supersymmetric attractors and also multiple attractors with one of the attractor points being 
supersymmetric in these four dimensional supergravity theories there is no known example where both the attractor 
points are supersymmetric  for these $N=2$ supergravity theories in four  dimensions \cite{Dominic:2014zia}. 
This suggests that, unlike the five dimensional case, the supersymmetric attractors might be unique in these
four dimensional supergravity theories. The present work aims to investigate this issue in detail. 

The plan of this paper is as follows. In the following section, we will briefly overview the $N=2$ supergravity theory.
In \S3 we will prove that the axion free attractors in four dimensions are unique. Subsequently, we will generalise
this result for axionic attractors. This will be followed by an explicit construction of all supersymmetric attractors
in a simple two-parameter model in \S4.  Finally, we will be summarise our results  in \S5.

\section{Overview}

The Lagrangian density for the bosonic part of the four dimensional $N=2$ supergravity theory coupled t
o $n$ vector multiplet, is given by
\begin{eqnarray}\label{sugra}
{\cal L} = - \frac{R}{2} +  g_{a\bar b} \partial_\mu x^a \partial_\nu \bar x^{\bar b} h^{\mu\nu} 
-  \mu_{\Lambda\Sigma} {\cal F}^\Lambda_{\mu\nu}{\cal F}^\Sigma_{\lambda\rho} 
h^{\mu\lambda}h^{\nu\rho} -   \nu_{\Lambda\Sigma} {\cal F}^\Lambda_{\mu\nu}
*{\cal F}^\Sigma_{\lambda\rho} h^{\mu\lambda}h^{\nu\rho} \ .
\end{eqnarray}
Here $h_{\mu\nu}$ is the space-time metric, $R$ is the corresponding Ricci scalar, $g_{ab}$ is the metric on the 
vector multiplet moduli space parameterized  by the corresponding $n$ complex scalar fields $x^a$ and 
$A_\mu^\Lambda$ are the $(n+1)$ gauge fields with corresponding field strength ${\cal F}^\Lambda_{\mu\nu}$. 
The gauge couplings $\mu_{\Lambda\Sigma}, \nu_{\Lambda\Sigma}$ and the moduli space metric $g_{a\bar b}$
are uniquely determined by the $N=2$ prepotential $F$.

We are interested in static, spherically symmetric configurations. The line element corresponding to the 
space time metric $h_{\mu\nu}$ in this case is given by
\begin{eqnarray}
ds^2 = e^{2U} dt^2 - e^{-2U}\gamma_{mn} dy^mdy^n \ .
\end{eqnarray}
The wrap factor $U$ depends only on the radial coordinate $r$. For extremal black holes, the metric of the 
spacial section $\gamma_{mn}$ must be identity. The equations of motion for these configurations simplifies 
and the system can now be described in terms of an effective one dimensional theory with a potential which 
is extremized  at the horizon.

For the $N=2$ Lagrangian (\ref{sugra}), the effective black hole potential takes the form \cite{Ferrara:1997tw}:
\begin{equation}
V=e^K \Big[g^{a \bar{b}} \nabla_a W \overline{\nabla_b W}+|W|^2\Big] \ . 
\end{equation}
Here $W$ and $K$ are respectively  the superpotential and the K\"ahler potential.  The superpotential $W$ is
related to the central charge by $Z = e^{K/2}W$. In terms of the dyonic charges $(q_\Lambda, p^\Lambda)$ 
and the prepotential $F$, the expression for $W$ is given by
\begin{eqnarray}
W = \sum_{\Lambda=0}^n (q_\Lambda X^\Lambda-p^\Lambda \partial_\Lambda F) \ , 
\end{eqnarray}
The symplectic sections $X^\Lambda$ are related to the physical scalar fields by $x^a = X^a/X^0$. 
The K\"ahler potential is given in terms of $F$ by the relation:
\begin{equation}
K =  -\log\Big[i\sum_{\Lambda=0}^{n} (\overline{X^\Lambda} \partial_\Lambda F 
- X^\Lambda \overline{\partial_\Lambda F})\Big] \ . 
\end{equation}
The covariant derivative is defined as $\nabla_a W = \partial_a W + \partial_a K W$. For supersymmetric 
attractors $\nabla_aW = 0$. In general, the attractor points are determined by $\partial_a V = 0$.

Throughout this paper, we will focus on the $N=2$ prepotential which is of the form
\begin{equation}\label{prepot}
 F=D_{abc}\frac{X^aX^bX^c}{X^0} \ . 
\end{equation}
The above prepotential appears as the leading term  in the compactification 
of type $IIA$ string theory on a Calabi-Yau manifold ${\cal M}$ in the large volume limit. In this case, $D_{abc}$ 
are the triple intersection numbers $D_{abc} = \int_{\cal M} \alpha_a\wedge\alpha_b\wedge\alpha_c$, where 
the two forms $\alpha_a$ form a basis of $H^2({\cal M},\mathbb{Z})$. In this paper, we will use string theory 
terminologies to describe various charge configurations irrespective of whether the coefficients $D_{abc}$
are actually associated with a Calabi-Yau compactification or not.

In the following we will describe some of the well known supersymmetric attractor solutions. For this purpose 
we need explicit expressions for the K\"ahler  and the superpotentials. The K\"ahler potential $K$ corresponding 
to the $N=2$ prepotential $F$ has the following simple form
\begin{eqnarray} \label{kahler}
K = -\log[-iD_{abc}(x^a-\bar{x}^a)(x^b-\bar{x}^b)(x^c-\bar{x}^c)] \ .
\end{eqnarray}
(Now on we set the gauge $X^0=1$ without any loss of generality and express our formulae in terms of 
the physical scalar fields $x^a$.) The superpotential depends on the specific charge configurations. In this 
paper we will mainly focus on $D0-D4$ and $D0-D4-D6$ configurations. For the $D0-D4$ configuration, 
the superpotential is given by
\begin{eqnarray} 
 W = q_0-3p^aD_{abc}x^bx^c \ , %+p^0D_{abc}x^ax^bx^c
\end{eqnarray}
whereas for the $D0-D4-D6$ configuration, we have 
\begin{eqnarray}
 W = q_0-3p^aD_{abc}x^bx^c+p^0D_{abc}x^ax^bx^c \ . 
\end{eqnarray}

These configurations possess well known supersymmetric attractor solutions \cite{Shmakova:1996nz}. 
For the $D0-D4$ configuration, we have
$$ \nabla_a W = - 6 D_{ab} x^b - \frac{3 M_a}{M} W. $$
From here onwards we use the standard notations\cite{Tripathy:2005qp}\ 
$D_{ab} = D_{abc} p^c, D_a = D_{ab}p^b, D = D_a p^a$, 
$ M_{ab} = D_{abc} (x^c - \bar x^c), M_a = M_{ab} (x^b - \bar x^b)$ and $M = M_a (x^a - \bar x^a)$.
(Note that $M_{a}$ is real where as $M_{ab}$ and $M$ are pure imaginary.)
Setting the ansatz, $x^a = p^a t$, we find 
$$ \nabla_a W =  - \frac{3 D_a}{2 t D } (q_0 + t^2 D ) \ , $$
and hence, $$x^a = i p^a \sqrt{\frac{q_0}{D}} \ , $$
for the supersymmetric $D0-D4$ configuration. The entropy of the corresponding supersymmetric 
black hole is $S = 2\pi \sqrt{q_0D}$.%\cite{Tripathy:2005qp}.

The solution can be generalised in a straightforward manner upon adding $D6$ branes. We find 
$$\nabla_a W = - 6 D_{ab} x^b + 3 p^0 D_{abc} x^bx^c - \frac{3 M_a}{M} W \ . $$
Setting the ansatz $x^a = p^a t$, we find the supersymmetric configuration corresponds to \cite{Shmakova:1996nz}
\begin{equation}\label{stdattr}
t = \frac{1}{2D}\Big( p^0 q_0 \pm i \sqrt{4 q_0 D - (p^0q_0)^2}\Big) \ . 
\end{equation}
The entropy for this configuration is 
$$ S = \pi  \sqrt{4 q_0 D - (p^0q_0)^2} \ . $$

\section{The general solution}

In this section, we will focus on the supersymmetric conditions more carefully and obtain the general
solution without assuming any specific ansatz. We will first focus on the $D0-D4$ configuration. Note
that, in this case the superpotential contains only even powers of $x^a$. Thus we can set 
the axionic parts of the scalar fields to zero: $x^a = i x_2^a$. The supersymmetry condition now becomes
\begin{equation}\label{susyeq}
M_{ab} p^b + \frac{M_a}{M}  (q_0 -\frac{ 3}{4} M_b p^b) = 0 \ . 
\end{equation}
Note that, for any configuration $x_2^a$ satisfying the above equation, we have $q_0 = -\frac{1}{4}M_a p^a$.
We can see this by multiplying by $(x^a-\bar{x}^a)$ and simplifying the above equation. Thus, we can further
simplify Eq.(\ref{susyeq}) by substituting $\frac{1}{4}M_a p^a =- q_0$ in it. We find
\begin{equation}
M_{ab} p^b  +4 q_0 \frac{M_a}{M} = 0 \ .
\end{equation}
Assuming the matrix $M_{ab}$ to be invertible,  we can rewrite the above equation as 
\begin{equation}\label{pa}
p^a =-8i q_0 \frac{x_2^a}{M} \ .
\end{equation}
This is a cubic equation in $x_2^a$. To solve it exactly, use the RHS of the above for $p^a$ in 
$D = D_{abc} p^ap^bp^c$ to rewrite it as $D = -64\frac{{q_0}^3}{M^2}$. Solving this for $M$ and 
substituting it in Eq.(\ref{pa}), we find $x^a = i p^a \sqrt{\nicefrac{q_0}{D}}$ as the most general 
axion free solution of the supersymmetric condition  (\ref{susyeq}).

We will now generalise this result in the presence of $D6$ branes. Note that in the presence of $D6$
branes it is no longer possible to set the axionic parts of the scalar fields to zero. We denote 
$x^a = x_1^a + i x_2^a$ and express the real and imaginary parts  the supersymmetric  condition 
$\nabla_aW=0$ as
\begin{eqnarray}
&& 4 M M_{ab}(p^b-p^0x^b_1) = 3 M_aM_b(p^b-p^0x^b_1) 
-4M_a(q_0-3D_{bc}x^b_1x^c_1+p^0D_{bcd}x^b_1x^c_1x^d_1) \ , \ \ \ 
\\
 &&8M D_{abc}x^b_1(2p^c-p^0x^c_1)-p^0MM_a = 12M_aM_{bc}x^b_1(2p^c-p^0x^c_1)\ .
\end{eqnarray}
For convenience we introduce $\omega^a=p^a-p^0 x^a_1$. Expressing the above equations in terms of 
$\omega^a$ and $x_2^a$, we find
\begin{eqnarray}
 && 4MM_{ab}\omega^b =3M_aM_b\omega^b-\frac{4M_a}{(p^0)^2}\big(q_0({p^0})^2-2D+3D_b\omega^b-D_{bcd}\omega^b\omega^c\omega^d\big) \ , \label{eq:A}
\\
&& \frac{8M}{p^0}(D_a-D_{abc}\omega^b\omega^c)-p^0MM_a = \frac{12M_a}{p^0}M_{bc}(p^bp^c-\omega^b\omega^c) \ . \label{eq:B}
\end{eqnarray}
We would like to find the most general solution of the above equations for the variables $\omega^a,x_2^a$. 
We first rewrite these equations in a simpler form so that it will be easier for us to solve them. Consider first
\eqref{eq:B}. Multiplying $(x^a-\bar{x}^a)$ on both side of of this equation and using the relation 
$D_a(x^a-\bar{x}^a) = M_{ab}p^ap^b$ we find
 \begin{equation}
 4 D_a( x^a-\bar{x}^a)+({p^0})^2M=4M_{ab}\omega^a\omega^b \ .
 \end{equation}
Using the above relation in \eqref{eq:B} we obtain
\begin{equation}
4 D_a+({p^0})^2M_a=4D_{abc}\omega^b\omega^c \ . \label{eq:1}
\end{equation}
We can similarly simplify \eqref{eq:A}. Multiplication of $( x^a-\bar{x}^a)$ on both sides of ~\eqref{eq:A} provides
\begin{equation}
 4\big(q_0({p^0})^2-2D+3D_a\omega^a-D_{abc}\omega^a\omega^b\omega^c\big)
 + ({p^0})^2 M_a\omega^a=0 \ . \label{eq:2}
\end{equation}
Putting back ~\eqref{eq:2} in ~\eqref{eq:A} we find
\begin{equation}
 MM_{ab}\omega^b=M_aM_b\omega^b\label{eq:3} \ . 
\end{equation}
Introducing  $\mu=(\nicefrac{ 2 i M_a\omega^a}{M})$  the above equation
 can be rewritten as $w^a= \mu x_2^a$.
 Substituting $\omega^a = \mu x_2^a$  in \eqref{eq:1} we get
% we see from \eqref{eq:6} 
$$D_a=-\frac{1}{4}({p^0}^2+\mu^2)M_a \ , $$ 
which implies
\begin{equation}\label{x2a}
 x_2^a=2i\frac{M^{ab}D_{bc}p^c}{{p^0}^2+\mu^2} \ .
\end{equation}
Defining  $${I^a}_b=2i\frac{M^{ac}D_{cb}}{\sqrt{{p^0}^2+\mu^2}} \ ,$$
 we can rewrite Eqs.\eqref{x2a} along with $\omega^a = \mu x_2^a$ as %and \eqref{eq:5} as 
\begin{eqnarray}
 w^a &=& \frac{\mu}{\sqrt{{p^0}^2+\mu^2}}{I^a}_bp^b \label{omegaa1} \ ,
 \\ \label{x2a1}
 x^a_2 &=& \frac{1}{\sqrt{{p^0}^2+\mu^2}}{I^a}_bp^b \ .
\end{eqnarray}
It can be shown that the  matrix ${I^a}_b$ is involutory: ${I^a}_b {I^b}_c = {\delta^a}_c$ and
it  satisfies the relation
\begin{eqnarray}
 D_{abc}{I^b}_e{I^c}_f = D_{aef}  \label{iabda} \ .
\end{eqnarray}
Using the explicit expressions for $\mu$ %as given in Eq.\eqref{eqmu} 
and after some simplifications, we can rewrite Eqs.\eqref{omegaa1} and \eqref{x2a1} in terms of the variables
$x_1^a,x_2^a$ as 
\begin{eqnarray}
x_1^a &=& \frac{1}{p^0}\bigg(p^a - \frac{D-\frac{1}{2}q_0{p^0}^2}{D_c{I^c}_dp^d}{I^a}_bp^b\bigg)\ ,\label{eq:g2} \\
x_2^a &=& \frac{1}{p^0}\bigg(\ 1-{\bigg(\frac{D-\frac{1}{2}q_0{p^0}^2}{D_c{I^c}_dp^d}\bigg)}^2\ \bigg)^{1/2}{I^a}_bp^b\ . \label{eq:g1}
\end{eqnarray}
This is the most general solution for the supersymmetry conditions \eqref{eq:A} and \eqref{eq:B}. Any involution
${I^a}_b$ satisfying the relation \eqref{iabda} will give us a new supersymmetric attractor. The standard solution
\eqref{stdattr} can be recovered by setting ${I^a}_b={\delta^a}_b$. We will have multiple attractors if there exists
nontrivial involutions satisfying \eqref{iabda} and if the moduli space metric as well as the gauge kinetic terms 
remain positive definite at more than one attractor points for the same charge configuration. 

For supersymmetric black holes the entropy is given by $S = \pi e^{K_0} |W_0|^2$, where $K_0$ and $W_0$
are the values of the K\"ahler and superpotential at the attractor point respectively. Substituting the value of 
$K_0$ and $W_0$ in the expression for entropy, we find
\begin{equation}
 S=\frac{\pi}{p^0}\sqrt{4 (D_a{I^a}_bp^b)^2-(2D-q_0{p^0}^2)^2} \ . 
\end{equation}

\section{An explicit example}

In the previous section we have derived the most general expression for the supersymmetric $D0-D4-D6$
attractors. They are given in terms of the involution ${I^a}_b$ satisfying the constraint \eqref{iabda}. In 
general it is not possible to solve \eqref{iabda} for arbitrary number of vector multiplets. Here we will 
consider the simplest case of a two-parameter model where this condition can be solved exactly to 
obtain new supersymmetric attractors.

A $2\times 2$ involution can be parametrised as 
\begin{equation}
{I^a}_b=\begin{pmatrix} u & v\\ w & -u
 \end{pmatrix}
\end{equation}
with $u^2+ v w=1$. To solve \eqref{iabda} for the two parameter case, we denote $D_{111} = a, D_{112} = b,
D_{122} = c$ and $D_{222} = d$. Further we use the notation 
$\mathcal{L}= a d-b c$, $\mathcal{M}=c^2-b d$ and $\mathcal{N}=b^2-a c$ for convenience.
Using $u^2+v w = 1$ we find two linearly independent equations from the condition \eqref{iabda}:
\begin{eqnarray}
&& a v - 2 b u - c w = 0 \ , \\
&& b v - 2 c u - d w = 0 \ . 
\end{eqnarray}
 It is straightforward to solve the above set of equations. For  $\mathcal{L}^2-4\mathcal{M}\mathcal{N}> 0$ 
 they admit a solution of the form:
 $$u=\frac{\mathcal{L}}{\sqrt{{\mathcal{L}}^2-4\mathcal{M}\mathcal{N}}} \ , v=\frac{-2\mathcal{M}}{\sqrt{{\mathcal{L}}^2-4\mathcal{M}\mathcal{N}}}\ ,  w=\frac{2\mathcal{N}}{\sqrt{{\mathcal{L}}^2-4\mathcal{M}\mathcal{N}}}$$
Thus we obtain a new $D0-D4-D6$ supersymmetric attractor in the two parameter case in addition to the
standard solution \eqref{stdattr}. Using the above solution for the involutory matrix ${I^a}_b$ we can obtain
explicit expressions for the vector multiplet moduli $x^1 = x^1_1 + i x^1_2$ and $x^2 = x^2_1 + i x^2_2$
(for easy reading we denote $\chi = D_a{I^a}_bp^b$ in the following):
\begin{eqnarray}
x^1_1 &=& \frac{1}{p^0}\bigg( p^1-\frac{(D-\frac{1}{2}q_0{p^0}^2)(\mathcal{L} p^1-2\mathcal{M} p^2)}{\chi\sqrt{{\mathcal{L}}^2-4\mathcal{M}\mathcal{N}}}\bigg) \ , \cr
x^1_2 &=& \frac{1}{p^0}\bigg(1-{\bigg(\frac{D-\frac{1}{2}q_0{p^0}^2}{\chi}\bigg)}^2\bigg)^{1/2} 
	    \frac{(\mathcal{L} p^1-2\mathcal{M} p^2)}{\sqrt{{\mathcal{L}}^2-4\mathcal{M}\mathcal{N}}}\label{eq:n1} \ , \cr
x^2_1 &=& \frac{1}{p^0}\bigg(p^2-\frac{(D-\frac{1}{2}q_0{p^0}^2)(2\mathcal{N} p^1-\mathcal{L} p^2)}{\chi\sqrt{{\mathcal{L}}^2-4\mathcal{M}\mathcal{N}}}\bigg) \ , \cr
x^2_2 &=& \frac{1}{p^0}\bigg(1-{\bigg(\frac{D-\frac{1}{2}q_0{p^0}^2}{\chi}\bigg)}^2\bigg)^{1/2} 
	    \frac{(2\mathcal{N} p^1-\mathcal{L} p^2)}{\sqrt{{\mathcal{L}}^2-4\mathcal{M}\mathcal{N}}}\label{eq:n2} \ .
\end{eqnarray}

Having obtained the above new configuration for the $D0-D4-D6$ attractors we would like to ask if it coexists 
with \eqref{stdattr} for the same set of charges. Both the solutions are well defined for 
$\mathcal{L}^2-4\mathcal{M}\mathcal{N}> 0$. However, this is not sufficient for the existence of the attractor
solution and we need to make sure that both the moduli space metric and the gauge kinetic terms are 
positive definite.

We will first consider the moduli space metric $g_{a\bar b} = \partial_a\partial{\bar b} K$. From the expression
for it K\"ahler potential \eqref{kahler} it is straightforward to find
\begin{equation}\label{mtrc}
 g_{a\bar{b}} = \frac{3}{M}\bigg(2M_{ab}-\frac{3}{M}M_aM_b\bigg) \ .
\end{equation}
At the attractor point \eqref{stdattr} it takes the form 
\begin{equation}
g_{a\bar{b}}=\frac{9}{q_0\big(4D-q_0{p^0}^2\big)}\big(D_aD_b-\frac{2}{3}DD_{ab}\big) \ ,
\end{equation}
where as for the new solution Eqs.\eqref{eq:g2} and \eqref{eq:g1} we have
 \begin{equation}
  g_{a\bar{b}}=\frac{9 {p^0}^2\chi}{4\Big(\chi^2-\big(D-\frac{1}{2}q_0{p^0}^2\big)^2\Big)}
  \bigg(D_aD_b-\frac{2}{3}\chi D_{abc}{I^c}_dp^d\bigg) \ . 
 \end{equation}

For the two parameter model it is straightforward to diagonalise both the metrics. The explicit expressions
for the eigenvalues are lengthy and we will not reproduce them here. For our purpose it will be sufficient
to consider the determinant of the metric. From \eqref{mtrc} we find 
\begin{eqnarray*}
\det{g } &=& (-1)^n\left(\frac{3}{M}\right)^{2n} \det{\bigg(M_a M_b - \frac{2 M}{3} M_{ab}\bigg)}\cr
		&=&   (-1)^n\left(\frac{3}{M}\right)^{2n}\bigg(\bigg(-\frac{2M}{3}\bigg)^{n-1}\bigg(
		\epsilon^{a_1a_2 \cdots a_n }M_1M_{a_1}M_{2a_2} \cdots M_{na_n}+ \cdots \cr
		&&+ \epsilon^{a_1a_2 \cdots a_n }M_{1a_1}M_{2a_2} \cdots 
		M_{(n-1)a_{n-1}} M_nM_{a_n}\bigg)   +\bigg(-\frac{2M}{3}\bigg)^n\det{M_{ab}}\bigg)
		%&=& 	 (-1)^n\left(\frac{3}{M}\right)^{2n} \bigg(M\bigg(-\frac{2M}{3}\bigg)^{n-1}+\bigg(-\frac{2M}{3}\bigg)^n\bigg)\det{M_{ab}}\cr
%&=& -3^n2^{(n-1)}\det{\bigg(\frac{M_{ab}}{M}\bigg)}
\end{eqnarray*}
Note that 
$\epsilon^{a_1a_2 \cdots a_n }M_1M_{a_1}M_{2a_2} \cdots M_{na_n} = 
\epsilon^{a_1a_2 \cdots a_n }M_1(x^{b_1} - \bar x^{b_1}) M_{a_1b_1}M_{2a_2} \cdots M_{na_n} 
 = M_1 (x^1-\bar x^1)  \det(M_{ab}) . $ 
 There are $n$ such terms and adding them all we get $M \det(M_{ab})$.
Thus,  the determinant of the moduli space metric 
is found to be $ -3^n2^{(n-1)}\det{\bigg(\frac{M_{ab}}{M}\bigg)}$.
Substituting the explicit solutions, we find, for \eqref{stdattr},
\begin{equation}
\det{g} = \frac{18D^2(\mathcal{N}{p^1}^2-\mathcal{L}p^1p^2+\mathcal{M}{p^2}^2)}{{q_0}^2\big(4D-q_0{p^0}^2\big)^2} \ ,
\end{equation}
where as, for Eqs.\eqref{eq:g2} and \eqref{eq:g1}
\begin{equation}
\det{g} = -\frac{18{p^0}^4\chi^2}{\big(4\chi^2-\big(2D-q_0{p^0}^2\big)^2\big)^2} 
 (\mathcal{N}{p^1}^2-\mathcal{L}p^1p^2+\mathcal{M}{p^2}^2)\ .
\end{equation}
From the above, we find that both the determinant are proportional to 
$ (\mathcal{N}{p^1}^2-\mathcal{L}p^1p^2+\mathcal{M}{p^2}^2)$ with the proportionality factor being positive 
for the first one where as negative for the second solution. Clearly, for a given set of charges, both the terms
can't be made positive simultaneously. Thus the moduli space metric become positive definite in mutually exclusive regions
of the charge lattice. The attractor solution becomes unique in each of these domains. For the attractor 
point \eqref{stdattr}, this domain is specified by $ (\mathcal{N}{p^1}^2-\mathcal{L}p^1p^2+\mathcal{M}{p^2}^2)>0$
where as for the solution \eqref{eq:n2} it is given by 
$ (\mathcal{N}{p^1}^2-\mathcal{L}p^1p^2+\mathcal{M}{p^2}^2) < 0$. We can explicitly verify that the 
eigenvalues become positive in these respective regions of the charge lattice. We have numerically 
verified that the gauge kinetic terms can also simultaneously be made positive definite by suitable 
choice of charges.

\section{Summary}

In this paper we have studied the uniqueness of supersymmetric attractors in $N=2$ supergravity theories in
four dimensions arising from type $IIA$ compactification on a Calabi-Yau manifold. We have proved the 
uniqueness for $D0-D4$ attractors. We found that the supersymmetry conditions admit more general 
solutions if we include $D6$ charges in addition. These solutions are determined by involutions which
satisfies certain constraints. For the two parameter model we can explicitly solve the constraint to find 
two independent solutions for the attractor equation. However, they exist in mutually exclusive domains 
of the charge lattice. Hence, the attractors are unique in the respective domains.

\end{document}